\RequirePackage{fix-cm}
\documentclass[twocolumn]{svjour3}          
\usepackage[a4paper, total={7.42in, 8in}]{geometry}

\smartqed  

\usepackage{cuted}
\usepackage{lipsum}

\usepackage{dcolumn}
\usepackage{bm}
\usepackage{multicol}
\usepackage{mwe}
\usepackage{graphicx}
\usepackage{multicol,lipsum}
\usepackage{amsmath}
\usepackage{dcolumn} 
\usepackage{enumitem}
\usepackage{graphicx}
\usepackage{dcolumn}
\usepackage{array,multirow}
\usepackage{bm}
\usepackage{amsmath}
\usepackage{float}
\usepackage{epstopdf}
\usepackage{amsfonts}
\usepackage{amssymb}
\usepackage{mathrsfs}
\usepackage{epsfig}
\usepackage{tabularx}
\usepackage{cite}
\usepackage[dvipsnames]{xcolor}
\RequirePackage[colorlinks,citecolor=blue,urlcolor=blue,linkcolor=blue]{hyperref}
\newcommand{\be}{\begin{equation}}
	\newcommand{\ee}{\end{equation}}
\newcommand{\beq}{\begin{eqnarray}}
	\newcommand{\eeq}{\end{eqnarray}}

\newcommand{\arcsinh}{\mathrm{arcsinh}}
\newcommand{\arccosh}{\mathrm{arccosh}}

\begin{document}

\title{Timelike geodesics for five-dimensional Schwarzschild and Reissner-Nordstr\"om  Anti-de Sitter black holes}

\author{P. A. Gonz\'{a}lez  \and Marco Olivares \and
        Yerko V\'{a}squez 
        \and J. R. Villanueva 
}

\institute{%
          P. A. Gonz\'{a}lez \and Marco Olivares \at  
              Facultad de Ingenier\'{i}a y Ciencias,\\
              Universidad Diego Portales,\\ 
              Avenida Ej\'{e}rcito Libertador 441, Santiago, Chile.
    \and Yerko V\'{a}squez \at
              Departamento de F\'{\i}sica, Facultad de Ciencias,\\ Universidad de La Serena,\\ 
              Avenida Cisternas 1200, La Serena, Chile.
    \and J. R. Villanueva \at
              Instituto de F\'isica y Astronom\'ia,\\ 
              Universidad de Valpara\'iso,\\
              Avenida Gran Breta\~na 1111, Valpara\'iso, Chile.\\
    \and P. A. Gonz\'alez \at
    \email{\textcolor{blue}{\href{mailto:pablo.gonzalez@udp.cl}{pablo.gonzalez@udp.cl} }}
    \and Marco Olivares \at
    \email{\textcolor{blue}{\href{mailto:marco.olivaresr@mail.udp.cl}{marco.olivaresr@mail.udp.cl} }}
    \and Yerko V\'asquez \at
    \email{\textcolor{blue}{\href{mailto:yvasquez@userena.cl}{yvasquez@userena.cl} }}
    \and J. R. Villanueva \at 
    \email{\textcolor{blue}{\href{mailto:jose.villanueva@uv.cl}{jose.villanueva@uv.cl} }} 
}

\date{Received: date / Accepted: date}

\maketitle

\begin{abstract}

The timelike structure of the five--dimensional 
Schwarzschild and Reissner-Nordstr\"om  Anti-de Sitter black holes is studied in detail. 
Different kinds of motion are allowed and studied by using an adequate effective potential. 
Then, by solving the corresponding equations of motion, several trajectories and orbits are described in terms of Weierstra\ss \, elliptic functions and elementary functions for neutral particles.

\end{abstract}

\tableofcontents


\section{Introduction}
\label{intro}

The study of the gravitation in higher dimensional space-times has been widely considered since the revolutionary ideas of Kaluza and Klein \cite{Kaluza21,Klein26}, where it was intended to unify electromagnetic theory with the also revolutionary Einstein's theory of general relativity, to the modern ideas of string theory, where the action contains terms that appear as corrections to the Einstein-Hilbert action \cite{Font:2005td, Chan:2000ms}.
In this sense, the vacuum solutions of general relativity for five dimensional, static, spherically symmetric black hole, known as the {\it Schwarzschild-Tangherlini} black holes solution, was obtained in Ref. \cite{Tangherlini:1963bw}, and represents the starting point for the study of more general five dimensional space-times. In particular, the Reissner-Nordstr\"om anti-de Sitter (RNAdS) black hole solution posses interesting features in the context of the AdS/CFT correspondence \cite{Maldacena:1997re, Witten:1998qj,Gubser:1998bc}, and other realizations \cite{Friess:2006kw,Saadat:2012zza,Villanueva:2013zta,Gonzalez:2020zfd}.
The properties of Reissner-Nordstr\"om black hole in $d$-dimensional anti-de Sitter space-time has been studied in Refs. \cite{Chamblin:1999tk,Chamblin:1999hg}, and some studies of geodesics of test particles in higher dimensional black hole spacetimes has been analyzed in Refs. \cite{Kovacs:1984qx,PhysRevLett.98.061102, Seahra:2001bx,Frolov:2003en,Hackmann:2008tu,Guha:2010dd, Gibbons:2011rh, Guha:2012ma, Kagramanova:2012hw, Chandler:2015aha,Gonzalez:2015qxc,Grunau:2017uzf, Kuniyal:2017dnu}.
Also, interesting aspects of the thermodynamic of higher 
dimensional black holes can be studied following the Refs. 
\cite{Fernandes,andre,kong}.


In this work we have extended the study of the geodesic structure of the five-dimensional RNAdS black hole that we carried out in \cite{Gonzalez:2020zfd}, where the null structure of such black hole was studied in detail, finding, among other things, new classes of orbits, genuine of this space-time, as the hippopede  of  Proclus  geodesic. It is worth to mention that the thermodynamics and the stability of the spacetime under consideration were studied via a thermodynamic point of view, and it was found special conditions on the black hole mass and the black hole charge where the black hole is in a stable phase \cite{Saadat:2012zza}. Also, the radial and circular motion of massive particles for this spacetime was analized in Ref. \cite{Guha:2010dd}, where the geodesics were studied from the point of view of the effective potential formalism and the dynamical systems approach. 
Finally, the shadow of  the five-dimensional RN AdS black hole  was studied in Ref. \cite{Mandal:2022oma}.

The manuscript is organized as follows: in Sec. \ref{GLBHS} we 
{ briefly review the main aspects of the spacetime under consideration}
present the bases of spacetime under consideration. Then, in Sec. \ref{nullgeod}, we { solve the (time-like) geodesic equation and study in detail the different orbits for neutral particles}. Finally, conclusions and final remarks are presented in Sec. \ref{conclusion}.

\section{Five-dimensional Schwarzschild and Reissner-Nordstr\"om anti-de Sitter black hole}
\label{GLBHS}

The five-dimensional RNAdS
black hole is a solution of the field equations that arise from the action \cite{Chamblin:1999hg}
\begin{equation}
    S=-\frac{1}{16\pi G_5}\int d^5x\sqrt{-g}(R-2\Lambda-F^2)\,,
\end{equation}
where $G_5$ is the 5-dimensional gravitational constant, $R$ is the Ricci scalar, $F^2$ represents the electromagnetic Lagrangian, $\Lambda=-6/\ell^2$ is the cosmological constant, and $\ell$ is the radius of AdS$_5$ space. The static, spherically symmetric metric that solves the field equations is given by
\begin{equation}\label{metr}
{\rm d}s^2=-f(r)\,{\rm d}t^2+\frac{1}{f(r)}{\rm d}r^2+r^2\,{\rm d}\Omega_{3}^2\,,
\end{equation} where the lapse function $f(r)$ is given by \footnote{The lapse function of the $(n+1)$-dimensional RNAdS spacetime is  given by 
\begin{equation}
    f(r)=1-\frac{m}{r^{n-2}}+\frac{q^2}{r^{2n-4}}+\frac{r^2}{\ell^2},
\end{equation}
where $m$, and $q$ are arbitrary constants. The parameter $m$ is related to the ADM mass $\mathcal{M}$ of the spacetime through
\begin{equation}
    \mathcal{M}=\frac{(n-1)\omega_{n-1}}{16\pi G}m\,,
\end{equation}
where $\omega_{n-1}$ is the volume of the unit $(n-1)$-sphere. On the other hand,  the parameter $q$ is related to the electric charge by
\begin{equation}
    \mathcal{Q}=\sqrt{2(n-1)(n-2)}\left(\frac{\omega_{n-1}}{8 \pi G}\right)q\,.
\end{equation}
In this work, we consider $n=4$, $m\rightarrow (2M)^2$ and $q^2\rightarrow Q^4$,
thereby, $M$ and $Q$ are related to the total mass $\mathcal{M}$ and the charge $\mathcal{Q}$ of the spacetime via the relations
\begin{equation}
    (2M)^2=\frac{16\pi G \mathcal{M}}{(n-1)\omega_{n-1}}\,,\, Q^2=\frac{8 \pi G \mathcal{Q}}{\sqrt{2(n-1)(n-2)}\omega_{n-1}}\,.
\end{equation}} 
\begin{equation}
\label{lapse} 
f(r)=1- \frac{4M^2}{r^2}+ \frac{Q^4}{r^4}+ \frac{r^2}{\ell^2} \,,
\end{equation}
and ${\rm d}\Omega_{3}^2={\rm d}\theta^2+\sin^2 \theta \,{\rm d}\phi^2+\sin^2 \theta \, \sin^2 \phi\, {\rm d}\psi^2 $ is the metric of the unit 3-sphere.

This spacetime allows an event horizon $r_+$, and a Cauchy horizon $r_{-}$, obtained from the  equation $f(r)=0$, 
which leads to the sixth degree polynomial  
\begin{equation}\label{p6}
P(r) \equiv r^6+\ell^2\,r^4-4 M^2 \ell^2\,r^2+\ell^2 Q^4=0\, .
\end{equation}
The change of variable $x=r^2-\ell^2 /3$ allows
to write Eq. (\ref{p6}) as
$P(x)=x^3-\alpha x+ \beta$,
where
\begin{equation}
\alpha=\ell^2\left(4M^2+ \frac{\ell^2}{3}\right) 
, \,\,\, \beta=\ell^2\left( Q^4+\frac{4M^2\ell^2}{3}+\frac{2\,\ell^4}{27}\right),
\end{equation}
 therefore, the Cauchy and the event horizons can be written as
\begin{equation}
r_{-}=\sqrt{x_0-\ell^2/3},\quad r_{+}=\sqrt{x_1-\ell^2/3}\,, 
\end{equation}
where
$x_{n}=Z\sin \left( \Theta+2n \pi /3  \right)$,
$Z=2 \sqrt{\alpha/3}$ and
\newline
$\Theta=\frac{1}{3} \arcsin \left( \frac{3 \sqrt{3} \beta}{2 \alpha^{3/2}}\right)$. The extremal black hole is characterized by the condition $r_{ext} =r_{+ }=r_{-}$, obtained from the relation 
\begin{equation}
\label{equ}
4\ell^4\left(4M^2-Q^2 \right) +8\ell^2M^2\left(32M^4-9Q^4 \right)-27Q^8 =0\,.     
\end{equation}

Note that when $Q=0$, the lapse function reduces to the five-dimensional Schwarzschild anti-de Sitter black hole, and the spacetime allows one horizon (the event horizon $r_+$) given by 
\begin{equation}
r_+=\ell\,\sinh \left[{1\over 2} \arcsinh\left({4M\over \ell} \right) \right]\,.  
\end{equation}


\section{The timelike structure for neutral particles}\label{nullgeod}
\begin{figure}[H]
	\begin{center}
		\includegraphics[width=60mm]{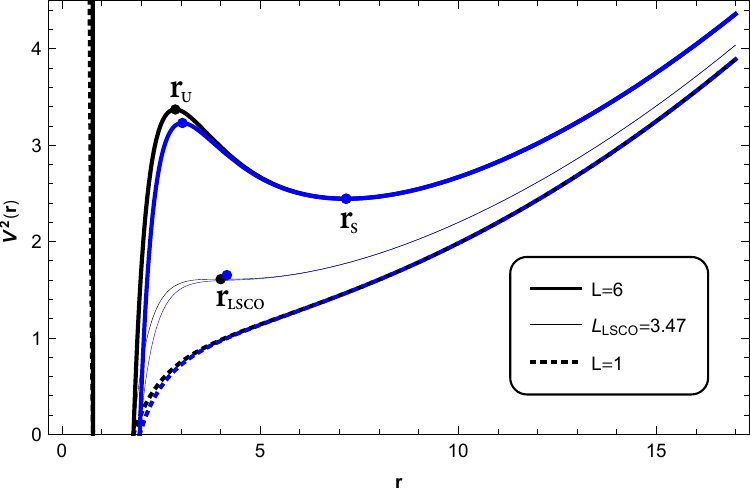}
	\end{center}
	\caption{The behaviour of the effective potential of neutral particles as a function of $r$ for different values of the angular momentum.
		Here, we have used the values  $M=1$, $\ell=10$, $Q=1.2$ (black lines), and $Q=0$ (blue lines). For $Q=1.2$, the unstable circular orbit occurs at $r_U=2.86$, the stable circular orbit at $r_S=7.18$, and the last stable circular orbit at $r_{\text{LSCO}}=4.01$, while that for $Q=0$ the unstable/stable/last stable circular orbits occurs at $r_U=3.04$, $r_S=7.17$ and $r_{\text{LSCO}}=4.16$, respectively. }
	\label{f3}
\end{figure}

\begin{figure}[H]
	\begin{center}
	\includegraphics[width=60mm]{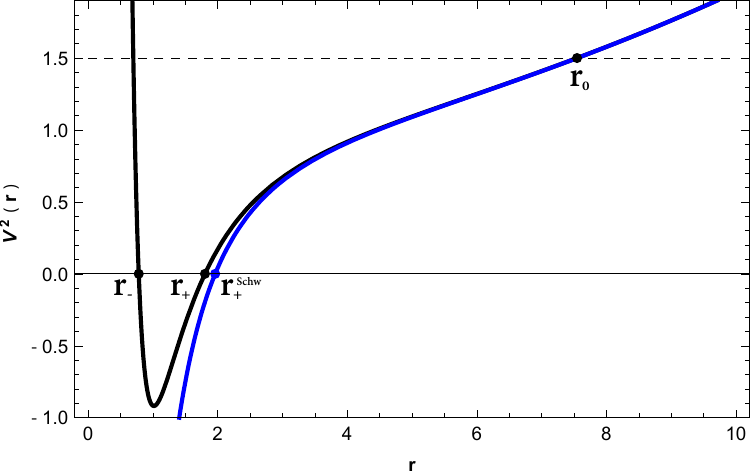}
	\end{center}
	\caption{The behaviour of the effective potential $V^2(r)$ for the motion of particles with $L=0$, $M=1$, and $\ell=10$. Black line for $Q=1.2$ ($r_0\approx 7.547$), and blue line for $Q=0$ ($r_0\approx 7.551$). Here, $r_-\approx 0.783$, $r_+\approx 1.805$, and $r_+^{\text{Schw}}\approx1.963$. }
\label{lapsus}
\end{figure}

In order to obtain a description of the allowed motion in the exterior spacetime of the black hole, we use the standard Lagrangian formalism \cite{Chandrasekhar:579245,Cruz:2004ts,Villanueva:2018kem}, so that, the corresponding Lagrangian associated with the line element (\ref{metr}) reads

\begin{equation}\label{Lag1}
\mathcal{L}=-\frac{f(r)\, \dot{t}^2}{2}+ \frac{\dot{r}^2}{2f(r)}+\frac{r^2}{2} \mathcal{L}_{\Omega}\,,  \end{equation}
where $\mathcal{L}_{\Omega}$ is the {\it angular Lagrangian}:
\begin{equation}\label{Lag2} 
\mathcal{L}_{\Omega}= \dot{\theta}^2+ \sin^2 \theta\,\dot{\phi}+\sin^2 \theta\,\sin^2 \phi\, \dot{\psi}^2\,.
\end{equation}
Here, the dot indicates differentiation with respect to the proper time $\tau$.  Since the Lagrangian (\ref{Lag1}) does not depend on the coordinates ($t,\psi$), they are {\it cyclic coordinates} and, therefore, the corresponding conjugate momenta $\pi_{q} = \partial \mathcal{L}/\partial \dot{q}$ is conserved along the geodesic. Explicitly, we have 
\begin{eqnarray}\label{pp1}
\pi_{t}&=& -f(r)\,,\,\, \dot{t}\equiv -E\,, \\\label{angcons} 
 \pi_{\psi}&=&r^2\sin^2 \theta \sin^2 \phi\, \dot{\psi}=L\,,
\end{eqnarray}where $E$ is a positive constant that depicts the time invariance of the Lagrangian, which cannot be associated with  energy because the spacetime defined by the line element (\ref{metr}) is not asymptotically flat, whereas the constant $L$ stands the conservation of angular momentum, under which it is established that the motion is performed in an invariant hyperplane. Here, we claim to study the motion in the invariant hyperplane $\theta=\phi = \pi/2$, so $\dot\theta =\dot\phi=0$ and, from Eq. (\ref{angcons}), we obtain 
\begin{equation} \dot{\psi}=\frac{L}{r^2}\,.
\label{eq1} 
\end{equation}
Therefore, using the fact that $2\mathcal{L}=-1$ for particles together with Eqs. (\ref{pp1}) and (\ref{eq1}), we obtain the following equations of motion 

\begin{eqnarray}
&&\left(\frac{{\rm d}r}{{\rm d} \lambda}\right)^{2}= E^2-V^2(r)\,,\\
\label{w.12}
&&\left(\frac{{\rm d} r}{{\rm d} t}\right)^{2}= {f^{\,2}(r)\over E^2}\left[E^2-V^2(r)\right]\,,\\
\label{w.13}
&&\left(\frac{{\rm d} r}{{\rm d} \psi}\right)^{2}=  {r^{\,4}\over L^2}\left[E^2-V^2(r)\right]\,,
\label{w.14}
\end{eqnarray}
where the effective potential $V^2(r)$ is defined by
\begin{eqnarray}\label{tl8}
\nonumber V^2(r)&\equiv& f(r)\left[1+\frac{L^{2}}{r^{2}}\right]=1-\frac{4M^2 }{r^2}+\frac{Q^{4}}{r^{4}}+\frac{r^{2}}{\ell^{2}}+\\
&& +\frac{L^{2}}{\ell^{2}}+\frac{L^{2}}{r^{2}}-\frac{4M^2 L^2}{r^4}+\frac{Q^{4} L^2}{r^{6}}\,.
\end{eqnarray}
The  behaviour of the effective potential is plotted in Fig. \ref{f3}, for test particles with angular momentum. Note that the following orbits are allowed: there are second kind trajectories for $0<L<L_{\text{LSCO}}$, there are last stable circular orbits ($\text{LSCO}$) and second kind trajectories for $L=L_{\text{LSCO}}$, and there are planetary orbits, circular orbits (stable and unstable) and second kind of trajectories for $L>L_{\text{LSCO}}$. While that for the motion with $L=0$, see Fig. \ref{lapsus}, there are second kind trajectories. Note that the effect of the black hole charge is the existence of a return point located at $r<r_-$.

In the next subsections, we will study the motion of neutral particles in five-dimensional backgrounds. Firstly, we will consider Schwarzschild black holes as uncharged spacetimes, and Reissner-Nordstr\"om black holes, as charged spacetimes.

\subsection{Five-dimensional Schwarzschild black hole}

In this section, firstly we will study the confined orbits, that is, circular and planetary orbits. Then, we will study the unconfined orbits, as second kind, and critical trajectories, and finally the motion of particles with vanishing angular momentum.  

\subsubsection{Circular orbits}

The existence of a maximum and a minimum in the potential corresponds to unstable (U)/stable(S) circular orbits, respectively. Thus, $V'=0$ yields
\begin{equation}\label{Pre}
   \tilde{P}(r)\equiv  r^6-\tilde{A} r^2 +\tilde{B}=0\,,
\end{equation}
where $\tilde{A}=\ell^2(L^2-4 M^2)$, and $\tilde{B}=8\ell^2 L^2 M^2$. Thereby, the particles can stay in unstable (stable) circular orbits of radius  $r_{U}$ ($r_{S}$), given by

\begin{equation}\label{rs} 
r_U=\left[\sqrt{\frac{4 \tilde{A}}{3}} \cos \left(\frac{1}{3} \cos ^{-1}\left(-12\tilde{B} \sqrt{\frac{3}{(4 \tilde{A})^3}} \right)+\frac{4 \pi }{3}\right)\right]^{1/2}\,, \\
\end{equation}
\begin{equation}\label{ru} 
r_S=\left[\sqrt{\frac{4 \tilde{A}}{3}} \cos \left(\frac{1}{3} \cos ^{-1}\left(-12\tilde{B} \sqrt{\frac{3}{(4 \tilde{A})^3}} \right)\right)\right]^{1/2}\,. 
\end{equation}

Now, in order to determine the analytical value of $L_{LSCO}$, where $r_U=r_S=r_{LSCO}$, we set to zero the discriminant $\Delta$ of Eq. (\ref{Pre}) given by  $\Delta = 4\tilde{A}^3-27\tilde{B}^2$, which yields

\begin{equation}
 L^6-L^4 \left(\frac{432 M^4}{\ell ^2}+12 M^2\right)+48 L^2 M^4-64 M^6   =0\,.
\end{equation}
Therefore,
\begin{equation}
    L_{LSCO}=\sqrt{z_0+\frac{144 M^4}{\ell ^2}+4 M^2}\,,
\end{equation}
where
\begin{equation}
  z_0=\sqrt{\frac{h_2}{3}} \cosh \left(\frac{1}{3} \cosh ^{-1}\left(3 \sqrt{\frac{3}{h_2^3}} h_3\right)\right) \,, 
\end{equation}
\begin{equation}
    h_2=\frac{4 \left(3456 M^6 \left(\ell ^2+18 M^2\right)\right)}{\ell ^4}\,,
\end{equation}
\begin{equation}
h_3=\frac{4 \left(6912 M^8 \left(\ell ^4+864 M^4+72 \ell ^2 M^2\right)\right)}{\ell ^6}\,.
\end{equation}

\begin{figure}[H]
	\begin{center}
		\includegraphics[width=60mm]{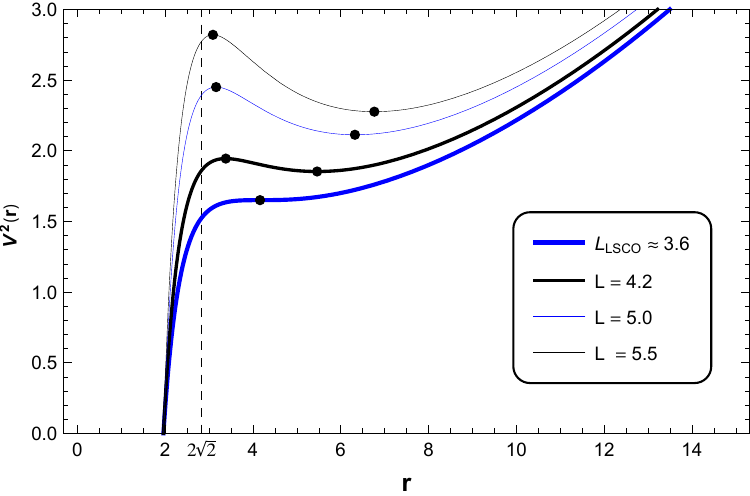}
	\end{center}
	\caption{The effective potential of Schwarzschild black hole for different values of the angular momentum $L$, with $M=1$, and $\ell=10$. Here, $r_{\text{LSCO}}=4.16$ for $L_{LSCO}=3.6$. Note that when the angular momentum increases, the radius of the unstable circular orbit tends to $2\sqrt{2}$, while that the radius of the stable circular orbit increases.}
	\label{fLSCO}
\end{figure}

In Fig. \ref{fLSCO}, we show the behaviour of the effective potential when the angular momentum increases. By using Eqs. (\ref{rs}), and (\ref{ru}) we obtain that $\lim_{L \to \infty}r_U=2\sqrt{2}M$, and $\lim_{L \to \infty}r_S=+\infty$, which is shown in Fig. \ref{fLSCO} for $M=1$.

On the other hand, the proper period in such circular orbit of radius $r_{c.o.}$ is
	\begin{equation}\label{p1}
	T_{\tau}=2\pi\,r_{c.o.}\ell\, \sqrt{\frac{ r_{c.o.}^2-8 M^2 }{r_{c.o.}^4+4M^2 \ell ^2  }}\,,
	\end{equation}
	and the coordinate period is
	\begin{equation}\label{p2}
	T_t=\frac{2\pi r^2_{c.o.}\ell}{\sqrt{r_{c.o.}^4+4\ell^2 M^2 }}\,.
	\end{equation}

Now, in order to obtain the epicycle frequency for the stable circular orbit we expand the effective potential around $r=r_S$, which yields
\begin{equation}\label{e17}
V(r)=V(r_S)+V'(r_S)(r-r_S)+{1\over2}V''(r_S)(r-r_S)^2+...\,,
\end{equation}
where$\,'$ means derivative with respect to the radial coordinate.
Obviously, in this orbits $V'(r_S)=0$. So, by defining the {\it smaller}
coordinate $x=r-r_S$, together with {\it the epicycle frequency}
$\kappa^2\equiv1/2V''(r_S)$ \cite{RamosCaro:2011wx},  we can rewrite the above equation as
\begin{equation}\label{e18}
V(x)\approx E_S^2+\kappa^2\,x^2\,,
\end{equation}
where ${E_S}$ is the energy of the particle at the stable circular orbit.
Also, it is forward to see that the test particles satisfy the harmonic equation of motion $\ddot{x}=-\kappa^2\,x$, and the epicycle frequency is given by
\begin{equation}
    \kappa^{2}=\frac{4 \left(-16 \ell^2 M^4-12 M^2 r_S^4+r_S^6\right)}{\ell ^2 r_S^4 \left(r_S^2-8 M^2\right)}\,.
\end{equation}

\subsubsection{Planetary orbits}

As mentioned, the planetary orbits are orbits of the first kind,  and they occur when $L>L_{LSCO}$, and the energy $E$ lies in the range $E_{S}<E<E_U$. 
Thus, we can rewrite
the characteristic polynomial as
\begin{equation}\label{PrS}
\left(\ell\, L \, \frac{{\rm d}r}{{\rm d}\psi}\right)^2=-\mathcal{P}(r)=-(r^6+\tilde{a}r^4+\tilde{b}r^2+\tilde{c})\,,
\end{equation}
where
$\tilde{a}=-\ell^2 (E^2-1 - L^2/\ell^2)$,
$\tilde{b}=-\ell^2(4 M^2-L^2)$, 
$\tilde{c}=-\ell^2  \left(4 L^2 M^2\right)$. 
So, $\mathcal{P}(r)=0$ allows three real positive roots, which we can identify as: $r_P$, it corresponds
to a {\it  periastro} distance,
$r_A$ is interpreted as a
{\it  apoastro} distance, and  $r_F$ is a turning point for the second kind trajectories. Thereby, the polynomial (\ref{PrS}) can be written as
\begin{equation}\label{c10}
\mathcal{P}(r)= \left(r^2-r_{A}^2\right)(r^2-r^2_P)(r^2-r_F^2)\,,
\end{equation}
where the return points are given by
\begin{eqnarray}\label{rpaf4} 
r_A&=&\left[  u_0-\frac{1}{3} \left(-\ell ^2 E ^2+\ell ^2+L^2\right) \right]^{1/2}\,, \\
r_P&=&\left[  u_2-\frac{1}{3} \left(-\ell ^2 E ^2+\ell ^2+L^2\right)\right]^{1/2}\,, \\ 
r_F&=&\left[  u_1-\frac{1}{3} \left(-\ell ^2 E ^2+\ell ^2+L^2\right)\right]^{1/2}\,, \\
\end{eqnarray}
and the constants are
\begin{eqnarray}\label{alfabetabar} 
u_n&=& \sqrt{{\tilde\eta_2\over 3}} \cos\left(\frac{1}{3}\arccos\left({3}\tilde\eta_3\sqrt{{3\over \tilde\eta_2^3}}  \right) +\frac{2 \pi  n}{3} \right)\,,\\
\nonumber\tilde\eta_2&=&\frac{4}{3} \left(\ell ^4 \left(E ^2-1\right)^2+L^4-\ell ^2 L^2 \left(2 E ^2+1\right)\right.\\
&&\left. +12 \ell ^2 M^2\right)\,,\\
\nonumber \tilde\eta_3&=&-4 \left(\frac{2}{27} \left(L^2-\ell ^2 \left(E ^2-1\right)\right)^3 -4 \ell ^2 L^2 M^2\right.\\
 && \left. -\frac{1}{3} \ell ^2 \left(L^2-4 M^2\right) \left(L^2-\ell ^2 \left(E ^2-1\right)\right)\right)\,.
\end{eqnarray}

Therefore, the solution of planetary geodesics is given by 
\begin{equation}
\label{rphi}
    r(\phi)=\frac{r_A}{\sqrt{4 \wp (\kappa_P \phi  + \varphi_P )+\kappa_2}}\,,
\end{equation}
where the integration constants are
\begin{equation}
   \varphi_ P=\wp ^{-1}\left(\frac{1-\kappa_2}{4}\right)\,,
\end{equation}
 \begin{equation}
     \kappa_2=\frac{1}{3} \left(1+\left(\frac{r_A}{r_F}\right)^2+\left(\frac{r_A}{r_P}\right)^2\right)\,,\,\, \kappa_p=\frac{2 r_P r_F}{\ell  L}\,,
 \end{equation}
and the Weierstra$\ss$ invariants are

\begin{eqnarray}
    \nonumber g_2&=&\frac{1}{12} \left(-\left(\frac{r_A}{r_F}\right)^2 \left(\frac{r_A}{r_P}\right)^2+\left(\frac{r_A}{r_F}\right)^4-\left(\frac{r_A}{r_F}\right)^2 \right.\\
    && \left. +\left(\frac{r_A}{r_P}\right)^4-\left(\frac{r_A}{r_P}\right)^2+1\right)\,,
\end{eqnarray}

\begin{eqnarray}
  \nonumber   g_3&=&\frac{1}{432} \left(\left(\frac{r_A}{r_F}\right)^2-2 \left(\frac{r_A}{r_P}\right)^2+1\right)\times \\
 \nonumber   && \left(2 \left(\frac{r_A}{r_F}\right)^2-\left(\frac{r_A}{r_P}\right)^2-1\right) \left(\left(\frac{r_A}{r_F}\right)^2+\left(\frac{r_A}{r_P}\right)^2-2\right)\,.\\
\end{eqnarray}

\begin{figure}[H]%
	\begin{center}
	\includegraphics[width=5cm]{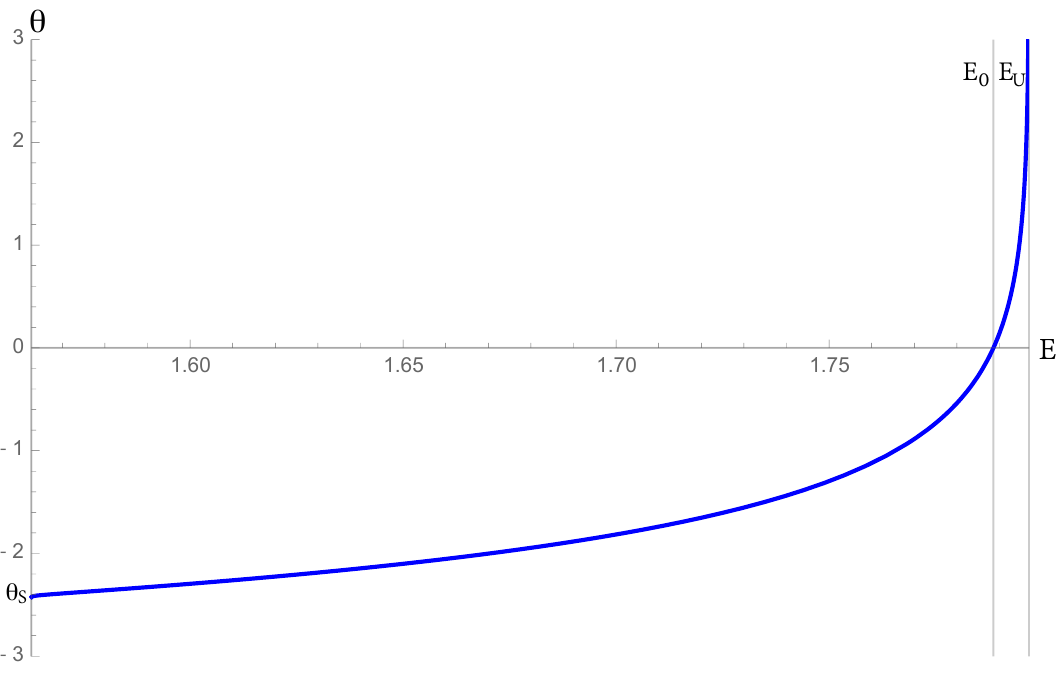}
	\end{center}
	\caption{The behaviour of the precession angle $\Theta$ as a function of the energy $E$, with  $M=1$, $\ell=10$, and $L=6$. Note that the precession angle vanishes when $E=E_0 \approx 1.789$, it tends to infinity when $E\rightarrow E_U \approx 1.797$, and $\Theta \rightarrow \Theta_S \approx -2.424$ when the energy $E \rightarrow E_S\approx 1.563$.
	}
	\label{fig:anguloalfa1S}
\end{figure}

\begin{figure}[H]
	\begin{center}
 \includegraphics[width=3cm]{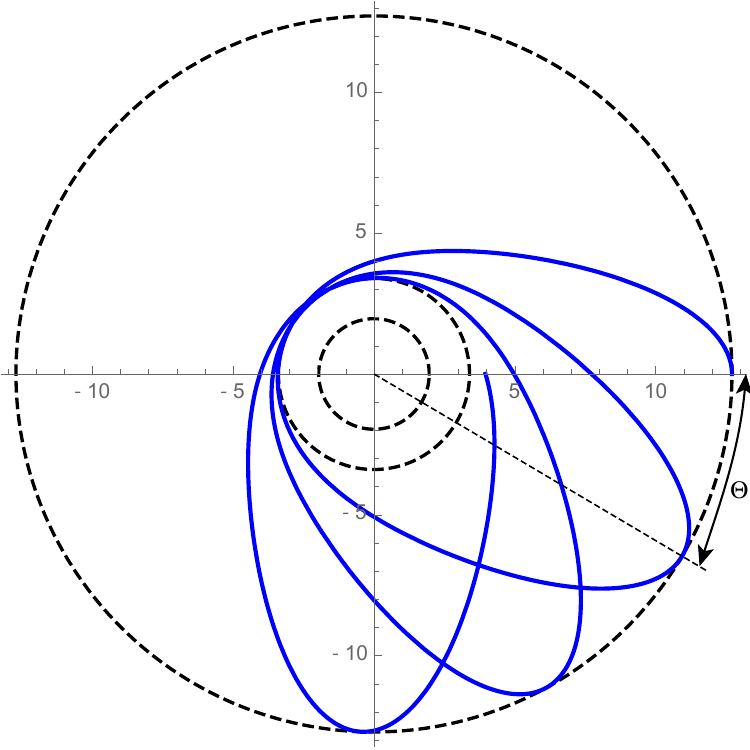}\\
		\includegraphics[width=3cm]{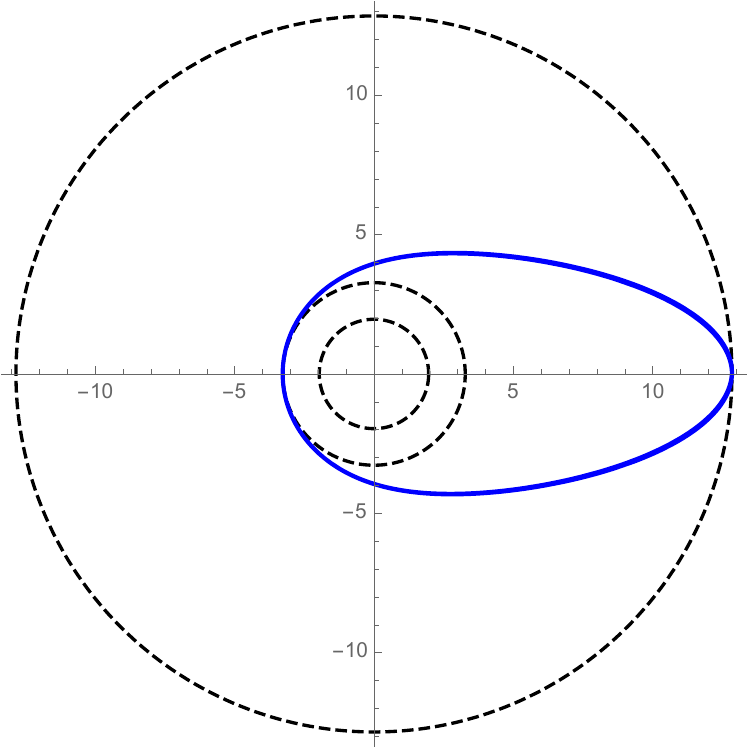}\\
		\includegraphics[width=3cm]{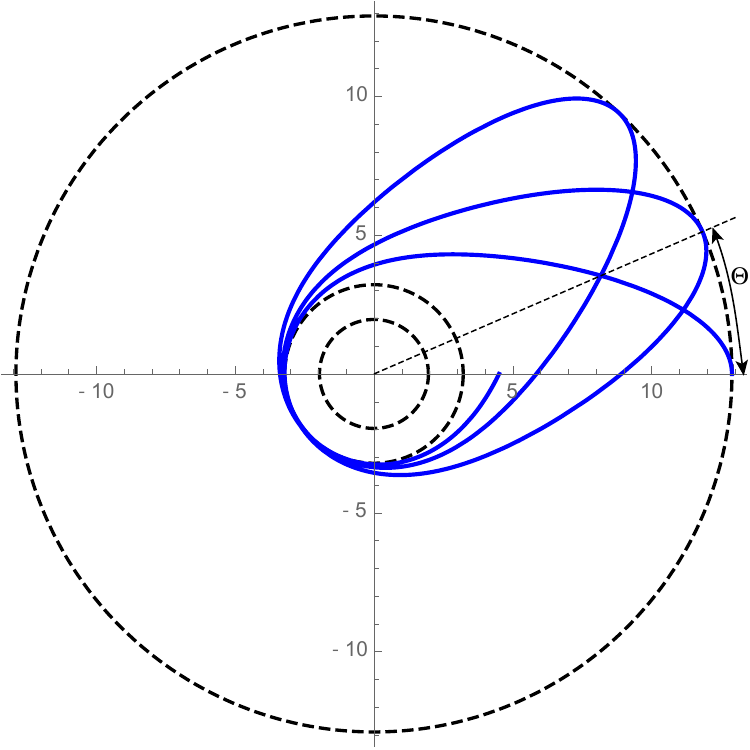}
		
	\end{center}
	\caption{The behaviour of second kind trajectories with  $M=1$, $\ell=10$, and $L=6$. Top panel corresponds to an orbit with negative precession $E \approx 1.74$, central panel without precession with $E\approx 1.790$, and bottom  panel for a positive precession with $E\approx 1.799$.}
	\label{Sfig:anguloalfa}
\end{figure}

Also, the solution (\ref{rphi}), allows us to determine the precession angle $\Theta$, by considering that it is given by $\Theta= 2\phi_P-2\pi$, where $\phi_P$ is the angle from the apoastro to the periastro. Thus, we obtain
\begin{equation}
\label{TS}
    \Theta =\frac{2}{\kappa_P} \left(\wp ^{-1}\left(\frac{r_A^2}{4 r_P^2}-\frac{\kappa_2}{4}\right)-\wp ^{-1}\left(\frac{1}{4}-\frac{\kappa_2}{4}\right)\right)-2 \pi\,.
\end{equation}
This an exact solution for the angle of precession, and it depends on the spacetime parameters and on the particle motion constants. So, according with Eq. (\ref{TS}),  in   Fig.~\ref{fig:anguloalfa1S}, we show the behaviour of the precession angle $\Theta$ as a function of the energy $E$, where the precession is negative for $E_s<E<E_0$, the precession is null for $E=E_0$, and it is positive for $E_0<E<E_U$, which we show in Fig. \ref{Sfig:anguloalfa}, via Eq. (\ref{rphi}).

\subsubsection{Second kind trajectories} 

\begin{figure}[H]
	\begin{center}
		\includegraphics[width=3cm]{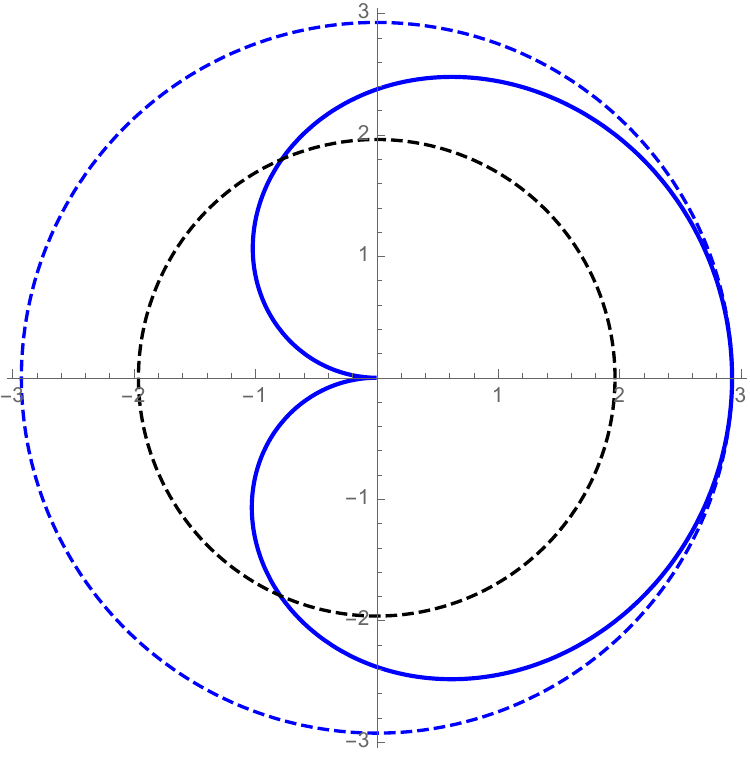}\\
		\includegraphics[width=3cm]{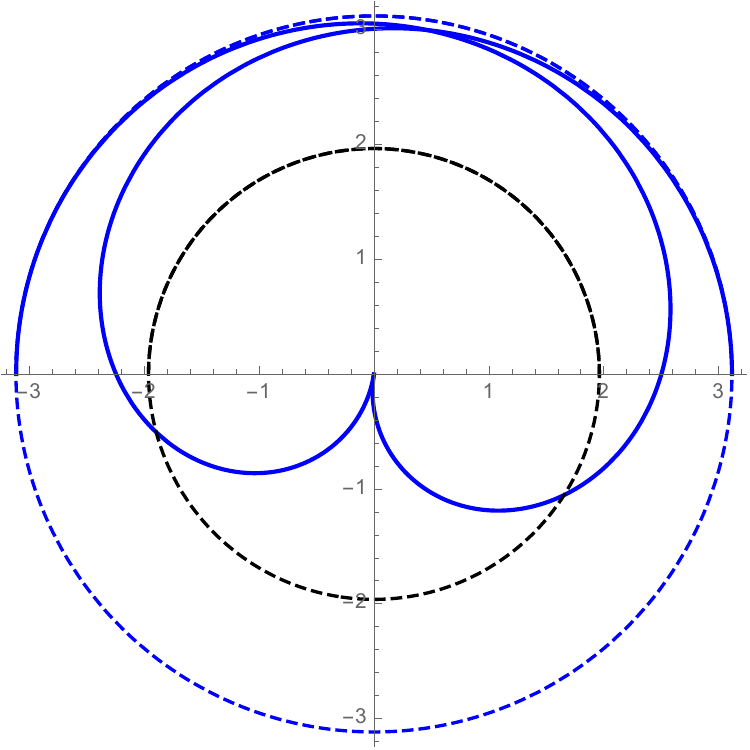}\\
		\includegraphics[width=3cm]{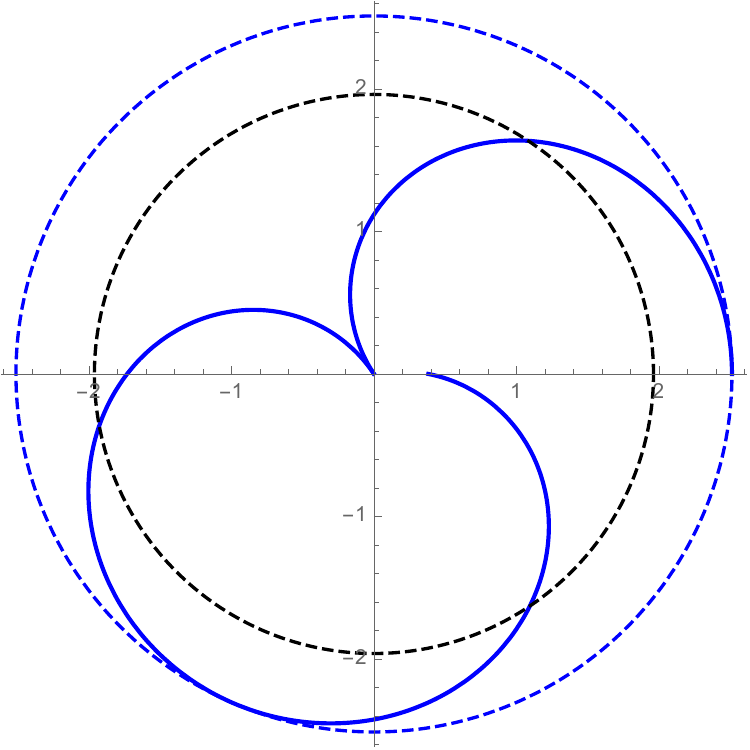}
	\end{center}
	\caption{The behaviour of second kind trajectories with  $M=1$, $\ell=10$, and $L=6$. Top panel corresponds to a orbit without precession with $E\approx 1.790$, central panel for a positive precession with $E\approx 1.799$, and bottom panel for negative precession $E=1.74$. The small circle represents to the event horizon, while that the bigger circle represents to the return point $r_F$. }
	\label{fig:anguloalfaS}
\end{figure}

The particles describe second kind trajectories when they start from a point of the spacetime and then they plunge into the horizon. In Fig. \ref{fig:anguloalfaS}, we show such trajectories. In the top panel, we observe that analytical extension of the geodesics show a similar trajectory to the cardioid \cite{Cruz:2004ts}. 

\begin{figure}[H]
	\begin{center}
		\includegraphics[width=3cm]{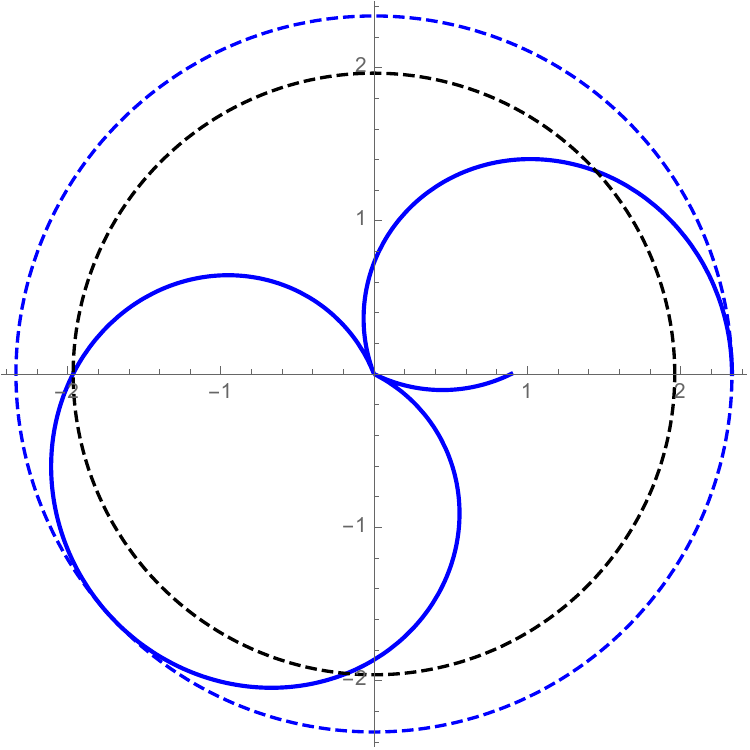}
	\end{center}
	\caption{The behaviour of a special second kind trajectories with  $M=1$, $\ell=10$, $L=6$, and $E=E_S=1.563$. The small circle represents to the event horizon $r_H\approx 1.963$, while that the bigger circle represents to the return point $r_F=2.5$. Here, $r_S\approx 6.325$.}
	\label{specialS}
\end{figure}

An analytical solution for second kind trajectories described by elementary functions can be found when $E=E_S$, and it is given by
\begin{equation}
    r_{2K}(\phi)=\frac{r_F r_S}{\sqrt{r_S^2+\left(r_S^2-r_F^2\right) \tan (\frac{r_S \sqrt{r_S^2-r_F^2}}{\ell L_P} \phi )^2}}\,.
\end{equation}
In Fig. \ref{specialS}, we plot the above solution and we can see a special second kind trajectory with a negative precession.

\subsubsection{Critical trajectories}

\begin{figure}[H]
	\begin{center}
		\includegraphics[width=3.0cm]{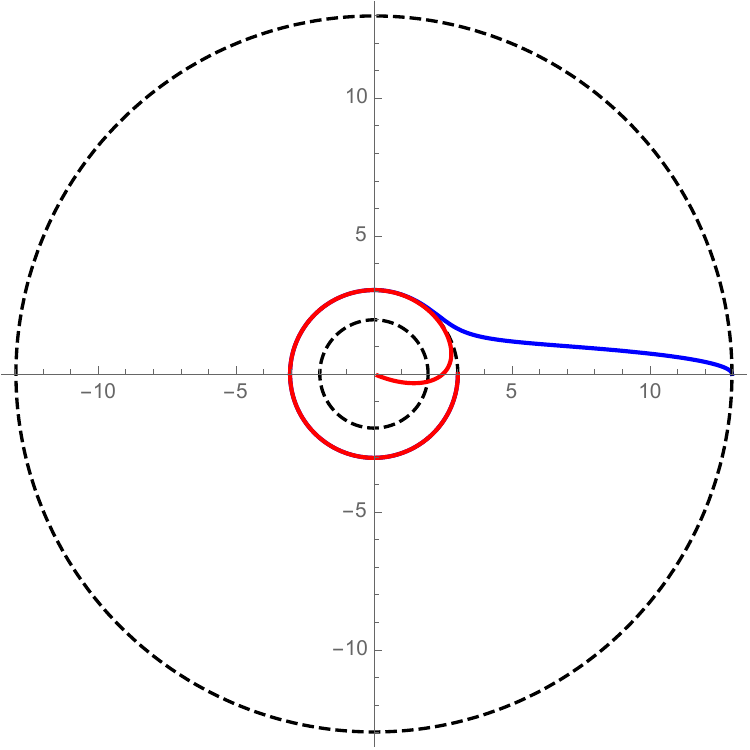}
	\end{center}
	\caption{The behaviour of critical trajectories with  $M=1$, $\ell=10$, and $L=6$. Blue line for CFK and red line for CSK trajectories with $E=E_U\approx 1.797$, and $r_U \approx 3.041$. The small dashed circle represents to the event horizon, while that the bigger circle represents to the return point $R_0$.
	}
	\label{fig:criticalS}
\end{figure}

This kind of motion is indeed ramified into two cases; critical trajectories of the first kind (CFK) in which the particles come from a distant position $r_0$ to $r_U$ and those of the second kind (CSK) where the particles start from an initial point near the horizon and can plunge into the horizon or they can tend to the stable circular orbit by spiraling. We obtain the following equations of motion for the aforementioned trajectories:

\begin{equation}
r_{CFK}(\phi)=\frac{R_0 r_U}{\sqrt{\left(R_0^2-r_U^2\right) \tanh ^2(k_1 \phi )+r_U^2}}\,,
\label{criticas1S}
\end{equation}
\begin{equation}
r_{CSK}(\phi)=\frac{R_0 r_U}{\sqrt{\left(R_0^2-r_U^2\right) \tanh ^2(k_1 \phi +\varphi_2)+r_U^2}}\,,
\label{criticas2S}
\end{equation}
where $\kappa_1=\frac{R_0 \sqrt{R_0^2-r_U^2}}{\ell  L}$, $\varphi_2=\tanh ^{-1}\left(\frac{r_U \sqrt{R_0^2-r_0^2}}{r_0 \sqrt{R_0^2-r_U^2}}\right)$,
and $R_0$ is $\left. r_{A} \right|_{E_U}$.
In Fig. \ref{fig:criticalS}, we show the behaviour of the CFK (blue line) and CSK (red line) trajectories, given by Eq. (\ref{criticas1S}) and Eq. (\ref{criticas2S}).

\subsubsection{Motion with $L=0$.}

For the motion of particles with $L=0$ the effective potential is $V^2=f(r)$. Consequently, the equations governing this kind of motion are
\begin{equation}
\frac{dr}{d\tau}=\pm\sqrt{E^2-f(r)}\,,
\label{mr.1}
\end{equation}
and
\begin{equation}
\frac{dr}{d t}=\pm\frac{f(r)}{E} \,\sqrt{E^2-f(r)}\,,
\label{mr.2}
\end{equation}
where the $+$ ($-$) sign  corresponds to particles  moving
toward the spatial infinite (event horizon). Now, we consider that the particles are placed at $r=\rho_0$
when $t=\tau=0$, which is also a turning point, and it is located at
\begin{equation}
\rho_{0}=\left[{\ell^2(1-E^2)\over 2}+\sqrt{\ell^4\left({1-E^2\over 2}\right)^2+4M^2\ell^2}\,\right]^{1\over 2}\,.
\end{equation}  
So, it is possible 
to obtain the proper time ($\tau$) for the motion of massive particles with $L=0$ in terms of the radial coordinate ($r$), resulting
\begin{equation}\label{t4}
\tau \left( r\right) ={\ell\over 2} \left[\sin^{-1}[\Xi(\rho_0)]-\sin^{-1}[\Xi(r)]\right]\,,
\end{equation}%
where
\begin{equation}
\Xi(r)={2\,r^2+\ell^2(1-E^2)\over\sqrt{16M^2\ell^2+\ell^4(1-E^2)^2}}\,,
\end{equation}
and the coordinate time ($t$) as function of $r$ yields
\begin{equation}\label{t5}
t\left( r\right) =\bar{A}_0\,\sum _{i=1}^{2}\bar{A}_i\,\left[\bar{F}_i(r)-\bar{F}_i(\rho_0) \right]\,,
\end{equation}
where the functions $\bar{F}_j(r)$ ($j=1,2$) are given explicitly by
\begin{eqnarray}
  	\label{F12}
\bar{F}_1(r)&=&\ln \left|\frac{\rho_0^2-r_+^2}{r^2-r_+^2}{Y_1(r)\over Y_1(\rho_0)}\right|\,,\\
\bar{F}_2(r)&=&\ln \left|\frac{\rho_0^2+r_+^2+\ell^2}{r^2+r_+^2+\ell^2}{Y_2(r)\over Y_2(\rho_0)}\right|\,,
  \end{eqnarray}
where
 \begin{eqnarray}
Y_1(r)&=&2k_1-k_2(r^2-r_+^2)+2\sqrt{k_1\,y_1(r)}\,,\\
Y_2(r)&=&2\bar{k}_1+\bar{k}_2(r^2+r_+^2+\ell^2)+2\sqrt{\bar{k}_1\,y_2(r)}\,.
  \end{eqnarray} 
   \begin{eqnarray}
y_1(r)&=&k_1-k_2(r^2-r_+^2)-(r^2-r_+^2)^2\,,\\
y_2(r)&=&\bar{k}_1-\bar{k}_2(r^2+r_+^2+\ell^2)-(r^2+r_+^2+\ell^2)^2\,.
  \end{eqnarray} 
with the corresponding constants,
\begin{equation}
    \bar{A}_0={E\,\ell^3\over 2(2r_+^2+\ell^2)}\,,\,\,
 \bar{A}_1={r_+^2\over \sqrt{k_1}}\,,\,\,
 \bar{A}_2={r_+^2+\ell^2\over \sqrt{\bar{k}_1}}\,,
\end{equation}   
   \begin{eqnarray}
 k_1&=&4M^2\ell^2-r_+^4-r_+^2\ell^2(1-E^2)\,,\\
 k_2&=&2r_+^2+\ell^2(1-E^2)\,,\\
 \bar{k}_1&=&4M^2\ell^2-(r_+^2+\ell^2)^2-(r_+^2+\ell^2)\ell^2(1-E^2)\,,\\
 \bar{k}_2&=&2(r_+^2+\ell^2)-\ell^2(1-E^2)\,.
  \end{eqnarray}
   
\begin{figure}[H]
	\begin{center}
	\includegraphics[width=50mm]{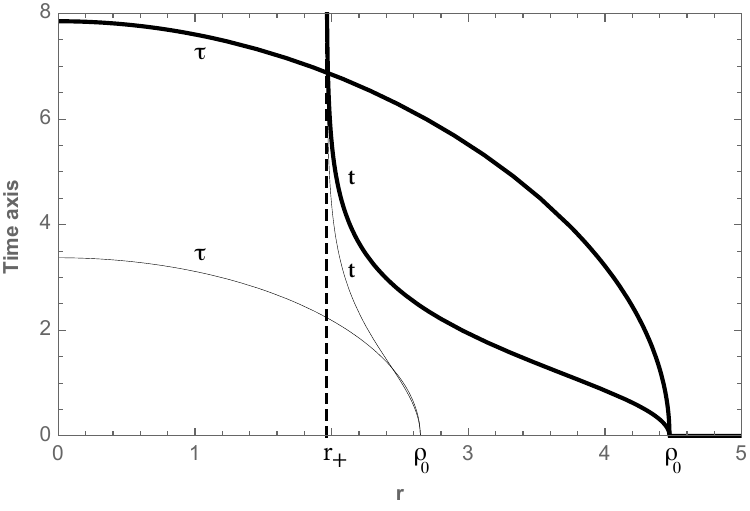}
	\end{center}
	\caption{Proper and coordinate time as a function of $r$ for the motion with $L=0$ of test particles. 
	Here, we have used the values $E\approx 0.707$, and $\rho_0\approx 2.649$ (thin curve), $E=1$, and  $\rho_0\approx 4.472$ (thick curve), with $M=1$, $\ell=10$. The dashed line corresponds to the event horizon ($r_+\approx 1.963$). }
	\label{f2S}
\end{figure}

Now, in order to visualize the behaviours of the proper and coordinate time, we plot in Fig \ref{f2S}, their behaviours via the solution (\ref{t4}), and (\ref{t5}), respectively. It is possible to observe that the particles moving to the event horizon, $r_+$, cross it in a finite proper time ($\tau$), but an external observer will see that the particles take an infinite (coordinate) time ($t$) to do it.

\subsection{Five-dimensional Reissner-Nordstr\"om black hole}

Now, for a charged spacetime, as previosly, firstly we will study the confined orbits, that is, circular and planetary orbits. Then, we will study the unconfined orbits, as the second kind, and critical trajectories, and finally the motion of particles with vanishing angular momentum.  

\subsubsection{Circular orbits.}

Similar to the previous background, we find two types of circular orbits.  Thus, the particles can stay in an unstable (stable) circular orbit of radius  $r_{U}$ ($r_{S}$), 
that can be obtained doing, $V'=0$. So,
\begin{equation}\label{PreRN}
   \bar{P}(r)\equiv  r^8+\bar{A} r^4 +\bar{B} r^2+\bar{C}=0\,,
\end{equation}
where
$\bar{A}=\ell^2(4 M^2-L^2)$, $\bar{B}=\ell^2  \left(8 L^2 M^2-2Q^4\right)$, and 
$\bar{C}=-3\ell^2 L^2 Q^4$. This eighth degree polynomial can be reduced to a fourth degree polynomial, where   
the radius are given by 
\begin{eqnarray}\label{rsRN} 
r_U&=&\left[  \bar{\alpha}+\sqrt{ \bar{\alpha}^2-\bar{\beta}} \right]^{1/2}\,, \\
r_S&=&\left[ \bar{\alpha}-\sqrt{ \bar{\alpha}^2-\bar{\beta}}\right]^{1/2}\,,
\end{eqnarray}
and the constants of the solution are 
\begin{equation}
    \bar{\alpha}=\left[  \bar{U}-{\bar{A}\over 6}\right]^{1/2}\,,\,\,
\bar{\beta}=2\bar{\alpha}^2 +{\bar{A}\over 2}+{\bar{B}\over 4\bar{\alpha}}\,,
\end{equation}
\begin{equation}
    \bar{U}= 2\sqrt{{\bar\eta_2\over 3}} \cosh\left(\frac{1}{3}\arccosh\left({3\over 2}\bar\eta_3\sqrt{{3\over \bar\eta_2^3}}  \right) \right)\,,
\end{equation}
\begin{equation}
    \bar\eta_2={\bar{A}^2\over 48}+{\bar{C}\over 4}\,,\,\,
\bar\eta_3={\bar{A}^3\over 864}+{\bar{B}^2\over 64}-{\bar{A} \bar{C}\over 24}\,.
\end{equation}

	The affine period in such circular orbit of radius $r_{c.o.}$ is
	\begin{equation}\label{p1}
	T_{\tau}=2\pi\,r_{c.o.}\ell\, \sqrt{\frac{ r_{c.o.}^4-8 M^2 r_{c.o.}^2+3 Q^4}{r_{c.o.}^6+4 \ell ^2 M^2 r_{c.o.}^2-2 \ell ^2 Q^4 }}\,,
	\end{equation}
	and the coordinate period is
	\begin{equation}\label{p2}
	T_t=\frac{2\pi\,r_{c.o.}^3\,\ell}{\sqrt{r_{c.o.}^6+4 \ell ^2 M^2 r_{c.o.}^2-2 \ell ^2 Q^4 }}\,.
	\end{equation}	
So, when the black hole charge is null, we recover the periods obtained for Schwarzschild black holes. \\

On the other hand, expanding the effective potential around
$r=r_S$, as in the previous case, it is possible to find the epicycle frequency for the stable circular orbit, which yields
\begin{equation}\label{e20}
\kappa^{2}= \frac{4 \left(r_S^{10}+k_0r_S^{8}+k_1r_S^{6}+k_2r_S^{4}+k_3r_S^{2}+k_4\right)}{\ell^2\, r_S^6 \left(r_S^4-8 M^2 r_S^2+3 Q^4\right)}\,,
\end{equation}
where $k_0=-12 M^2$, $k_1=6 Q^4$, $k_2=\ell^2\left( Q^4-16 M^4\right)$,  
$k_3=12 \ell^2 M^2 Q^4$, and  $k_4=-3 \ell ^2 Q^8$.
Notice that $\kappa \rightarrow \kappa_{Schw}$
when $Q\rightarrow 0$, where $\kappa_{Schw}$
is the epicycle frequency in the Schwarzschild case.\\

\subsubsection{Planetary orbit.}

It is possible to find confined orbits of first kind, for $L>L_{LSCO}$, for a charged background, as for Schwarzschild black holes. However, in this case an analytical expression for $L_{LSCO}$ can not be obtained. The characteristic polynomial can be written as  
\begin{equation}\label{Pr}
\left(\ell\, L \, r \frac{{\rm d}r}{{\rm d}\psi}\right)^2=-\mathcal{P}(r)=-(r^8+\tilde{a}r^6+\tilde{b}r^4+\tilde{c}r^2+\tilde{d})\,,
\end{equation}
where
$\tilde{a}=-\ell^2 (E^2-1 - L^2/\ell^2)$, 
$\tilde{b}=-\ell^2(4 M^2-L^2)$,  
$\tilde{c}=-\ell^2  \left(4 L^2 M^2-Q^4\right)$, and
$\tilde{d}=\ell^2 L^2 Q^4$.
Thus, $\mathcal{P}(r)=0$ allows four real roots, which we can identify as; $r_P$, corresponds
to a {\it  periastro} distance, whereas
$r_A$  
which is interpreted as a
{\it  apoastro} distance,  $r_F$ is a turning point for the trajectories of the second kind,
 and $r_4$  is a turning point smaller than the horizon $r_-$, that does not appear in the uncharged background. So, it is possible to write the characteristic polynomial (\ref{Pr}) as
\begin{equation}\label{c10}
\mathcal{P}(r)= \left(r^2-r_{A}^2\right)(r^2-r^2_P)(r^2-r_F^2)(r^2-r_4^2)\,,
\end{equation}
where
\begin{eqnarray}\label{rpaf4} 
r_A&=&\left[   \tilde{\alpha}+\sqrt{   \tilde{\alpha}^2- \tilde{\beta}} -{\tilde{a}\over 4}\right]^{1/2}\,, \\
r_P&=&\left[   \tilde{\alpha}-\sqrt{   \tilde{\alpha}^2- \tilde{\beta}} -{\tilde{a}\over 4}\right]^{1/2}\,, \\ 
r_F&=&\left[  - \tilde{\alpha}+\sqrt{   \tilde{\alpha}^2- \tilde{\gamma}} -{\tilde{a}\over 4}\right]^{1/2}\,, \\
r_4&=&\left[   -\tilde{\alpha}-\sqrt{   \tilde{\alpha}^2- \tilde{\gamma}} -{\tilde{a}\over 4}\right]^{1/2}\,.
\end{eqnarray}
and the constants are 
\begin{equation}
    \tilde{\alpha}=\left[  \tilde{U}-{\tilde{A}\over 6}\right]^{1/2}\,,\,\, 
\tilde{\beta}=2\tilde{\alpha}^2 +{\tilde{A}\over 2}+{\tilde{B}\over 4\tilde{\alpha}}\,,
\end{equation}
\begin{equation}
    \tilde{U}= 2\sqrt{{\tilde\eta_2\over 3}} \cosh\left(\frac{1}{3}\arccosh\left({3\over 2}\tilde\eta_3\sqrt{{3\over \tilde\eta_2^3}}  \right) \right)\,,
\end{equation}
\begin{equation}
    \tilde\eta_2={\tilde{A}^2\over 48}+{\tilde{C}\over 4}\,\,, \text{and}\,\,
\tilde\eta_3={\tilde{A}^3\over 864}+{\tilde{B}^2\over 64}-{\tilde{A} \tilde{C}\over 24}\,,
\end{equation}
with $\Tilde{A}=\tilde{b}-\frac{3\tilde{a}^2}{8}$, $\Tilde{B}=\tilde{c}+\frac{\tilde{a}^3}{8}-\frac{\tilde{a}\tilde{b}}{2}$,
and $\Tilde{C}=\tilde{d}+\frac{\tilde{a}^2\tilde{b}}{16}-\frac{3\tilde{a}^4}{256}-\frac{\tilde{a}\tilde{c}}{4}$.\\

Therefore, the solution for the planetary geodesics is given by 
\begin{equation}
\label{rphiRN}
    r(\phi)=r_A \sqrt{1-\frac{1}{\kappa_1+4 \wp (\kappa_p\phi)}}\,,
\end{equation}
where the integration constants are 
 
 \begin{equation}
     \kappa_1=\frac{1}{3} (Z_4+Z_F+Z_P)\,,
 \end{equation}

 \begin{equation}
     \kappa_p=\frac{2 \sqrt{\left(r_A^2-r_4^2\right) \left(r_A^2-r_F^2\right) \left(r_A^2-r_P^2\right)}}{  \ell L r_A}\,,
 \end{equation}
and the Weierstra$\ss$ invariants

\begin{equation}
    g_2=\frac{1}{12} \left(Z_4^2-Z_4 Z_F-Z_4 Z_P+Z_F^2-Z_F Z_P+Z_P^2\right)\,,
\end{equation}

\begin{equation}
    g_3=\frac{1}{432} (Z_4+Z_F-2 Z_P) (2 Z_4-Z_F-Z_P) (Z_4-2 Z_F Z_P)\,,
\end{equation}
whit

\begin{eqnarray}
     Z_4=\frac{r_A^2}{r_A^2-r_4^2}\,\,\,, Z_F=\frac{r_A^2}{r_A^2-r_F^2}\,\,\,, \text{and}\,\, 
     Z_P=\frac{r_A^2}{r_A^2-r_P^2}\,.
\end{eqnarray}\\

The solution (\ref{rphiRN}) allows us to determine the precession angle $\Phi$, by considering that it is given by $\Phi= 2\phi_P-2\pi$, where $\phi_P$ is the angle from the apoastro to the periastro. Thus, we obtain

\begin{equation}
\label{TRN}
    \Phi=\frac{2}{\kappa_P}\wp^{-1} \left[ \frac{r_A^2}{4 \left(r_A^2-r_P^2\right)}-\frac{\kappa_1}{4}\right]-2\pi\,.
\end{equation}

\begin{figure}[H]
	\begin{center}
	\includegraphics[width=5cm]{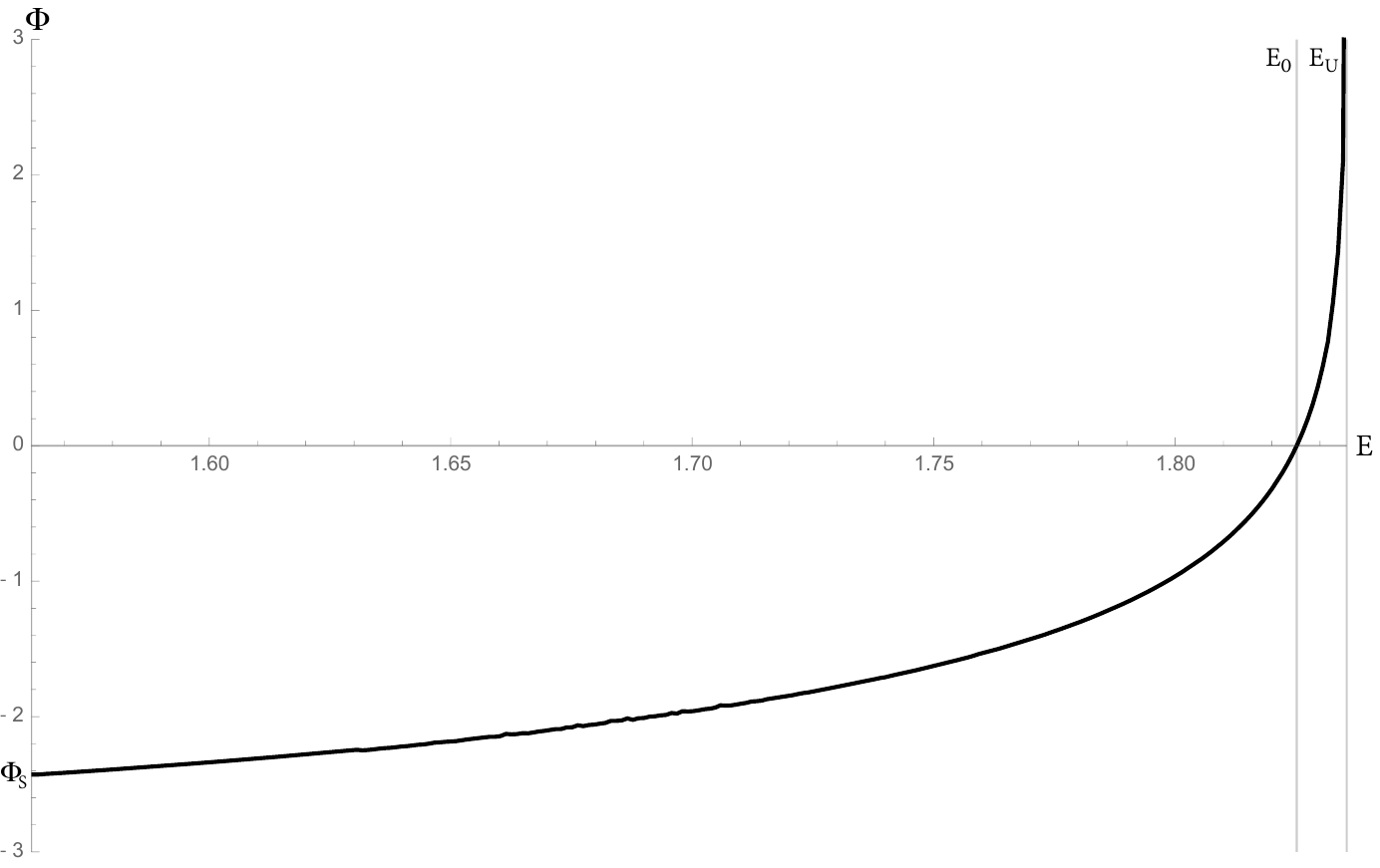}
	\end{center}
	\caption{The behaviour of the precession angle $\phi$ as a function of the energy $E$, with  $M=1$, $\ell=10$, $Q=1.2$, and $L=6$. Note that the precession angle vanishes when $E=1.825$, it tends to infinity when $E\rightarrow E_U \approx 1.836$, and $\Phi \rightarrow \Phi_S  \approx -2.428$ when the energy $E \rightarrow E_S \approx 1.563$.}
	\label{fig:anguloalfa1}
\end{figure}

\begin{figure}[H]
	\begin{center}
		\includegraphics[width=3.2cm]{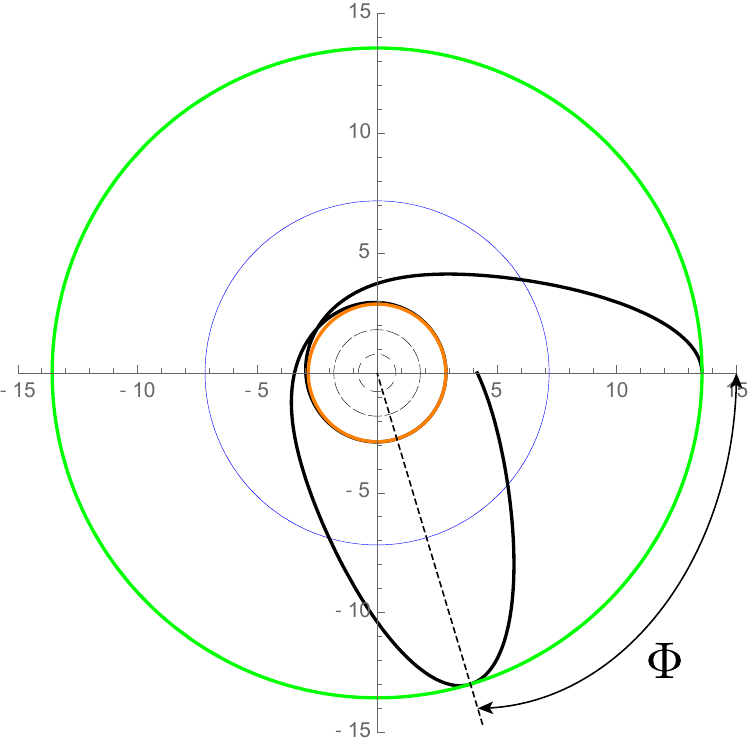}\\
        \includegraphics[width=3.2cm]{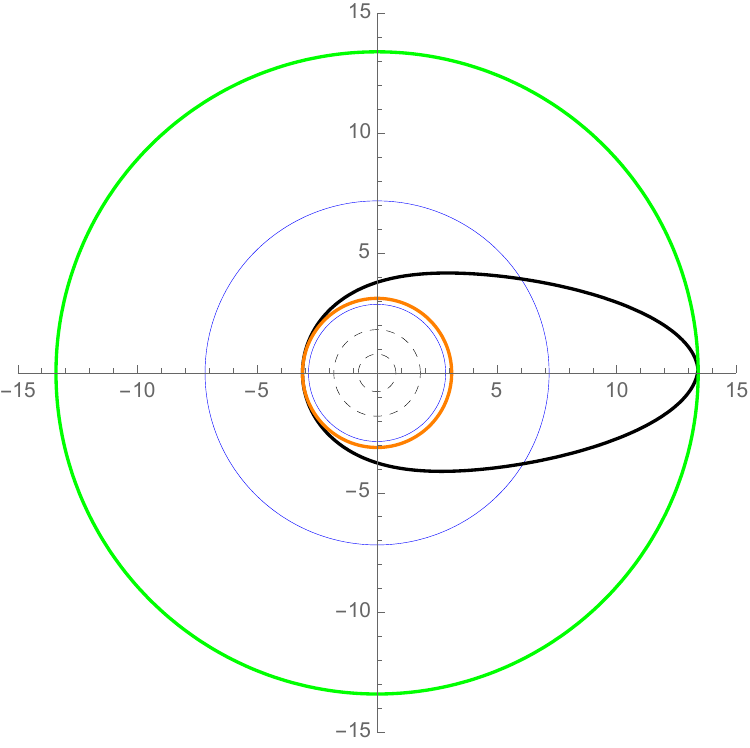}\\
		\includegraphics[width=3.2cm]{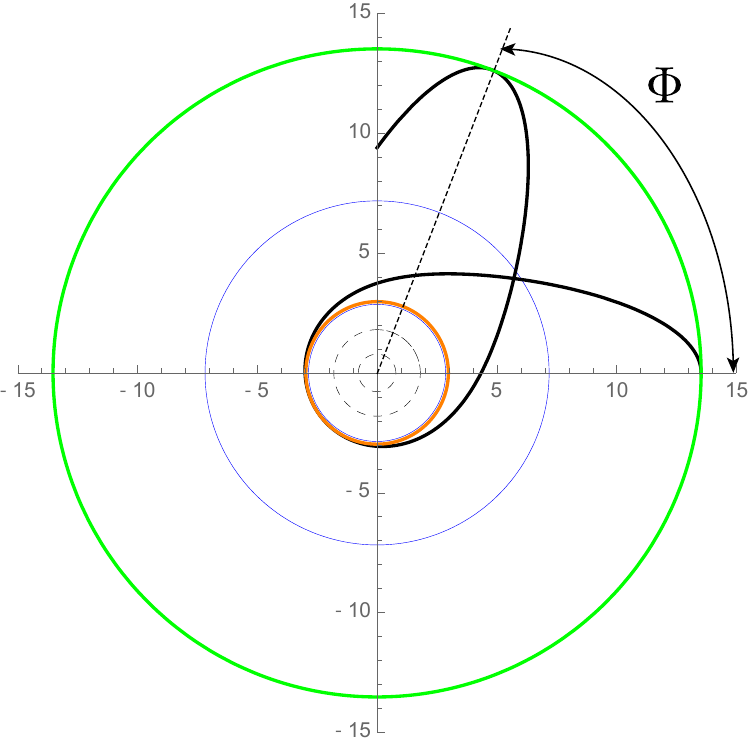}
		
	\end{center}
	\caption{The behaviour of the planetary orbits with  $M=1$, $\ell=10$, $Q=1.2$, and $L=6$. Top panel corresponds to an orbit with negative precession $E=1.835$, central panel without precession with $E=1.825$, and bottom panel for a positive precession with $E=1.833$. The green circle represents to the
	apoastro distance $r_A$, the  blue circle represents to the stable circular orbit, the orange circle represents to the periastro distance $r_P$, the small blue circle represents to the unstable circular orbit $r_U$, and the small dashed circle and the dashed circle corresponds to the Cauchy horizon $r_-$ and the event horizon $r_+$, respectively. }
	\label{fig:anguloalfaRN}
\end{figure}

This is an exact solution for the angle of precession, and it depends on the spacetime parameters and the particle motion constants. So, according with Eq. (\ref{TRN}), we show the behaviour of the precession angle $\Phi$ as a function of the energy $E$, in   Fig.~\ref{fig:anguloalfa1}, where for $E_s<E<E_0$ the precession is negative, for $E=E_0$ the precession is null and for $E_0<E<E_U$ it is positive, which we show in Fig. \ref{fig:anguloalfaRN}, via Eq. (\ref{rphiRN}).

\subsubsection{Second kind trajectories} 

\begin{figure}[H]
	\begin{center}
		\includegraphics[width=3cm]{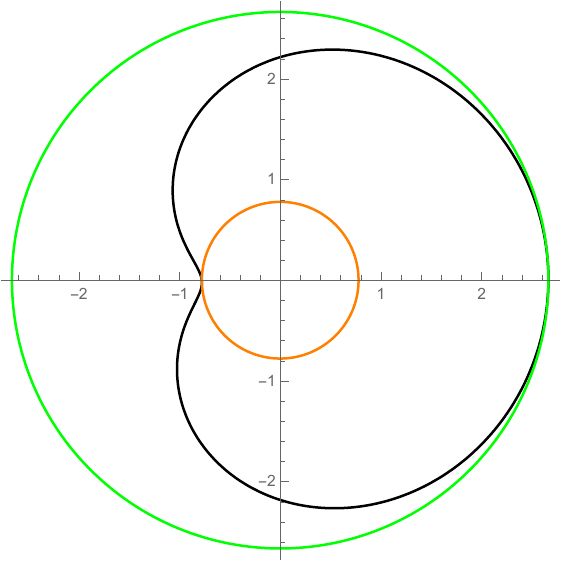}\\
		\includegraphics[width=3cm]{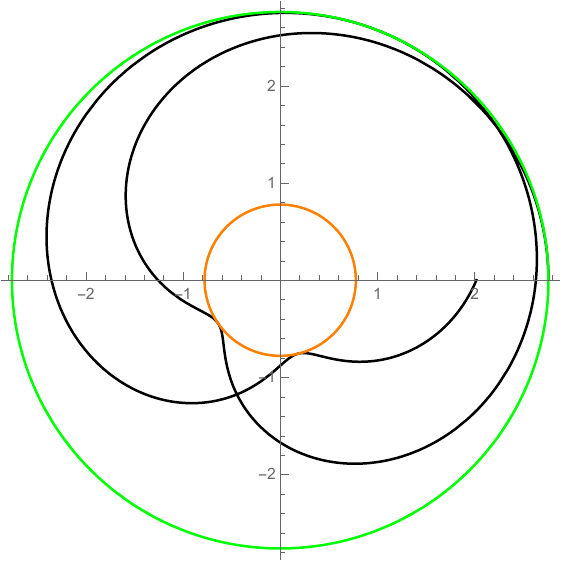}\\
		\includegraphics[width=3cm]{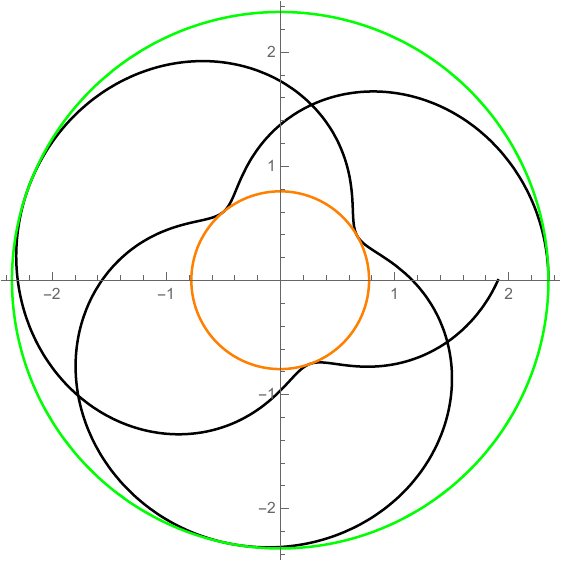}
	\end{center}
	\caption{The behaviour of second kind trajectories with  $M=1$, $\ell=10$, $Q=1.2$, and $L=6$. Top panel corresponds to an orbit without precession with $E\approx 1.825$, central panel for a positive precession with $E \approx 1.833$, and bottom panel for negative precession $E\approx 1.732$. The orange circle represents a return point ($r_4$)  inside the Cauchy horizon, and the green circle represents the return point ($r_F$) outside  the event horizon.}
	\label{RNfig:anguloalfa}
\end{figure}

The particles describe second kind trajectories when they start from a point of the spacetime and then they plunge into the horizon. In Fig. \ref{RNfig:anguloalfa} we show such trajectories with vanishing and non vanishing precession. In the top panel, we observe that the analytical extension of the geodesics shows a similar trajectory to the Limaçon of Pascal \cite{Villanueva:2013zta}. 

\begin{figure}[H]
	\begin{center}
		\includegraphics[width=4cm]{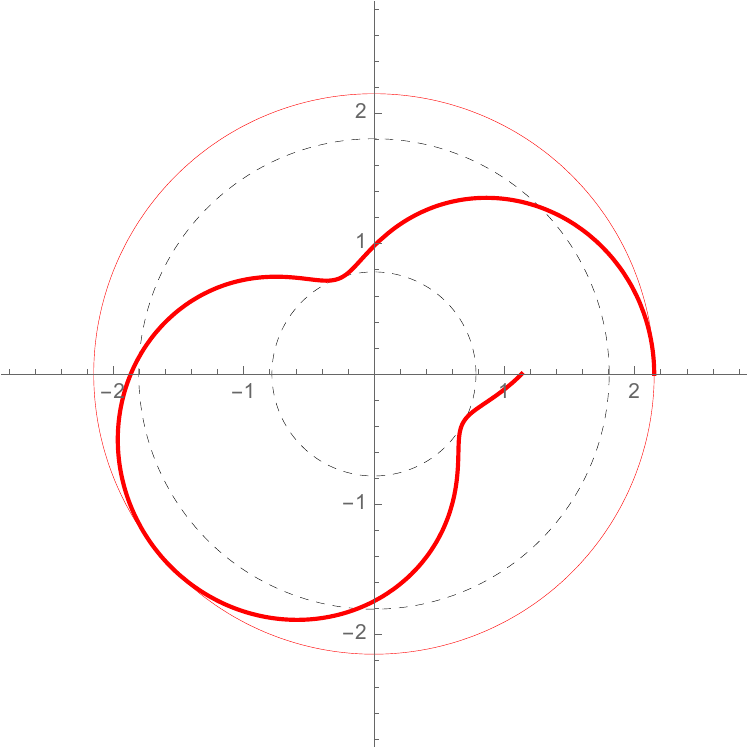}
	\end{center}
	\caption{The behaviour of a special second kind trajectories with  $M=1$, $\ell=10$, $Q=1.2$, $L=6$, and $E=E_S=1.563$. The small circle corresponds to the return point $r_4$, the medium circle represent to the event horizon $r_+$, and the bigger circle represents the return point $r_F$.}
	\label{SRNfig:anguloalfa}
\end{figure}

An analytical solution for second kind trajectories described by elementary functions can be found when $E=E_S$, and it is given by
\begin{equation}
    r_{2K}(\phi)=\sqrt{r_S^2-\frac{2 \left(r_S^2-r_4^2\right) \left(r_S^2-r_F^2\right)}{\left(-r_4^2-r_F^2+2 r_S^2\right)-\left(r_F^2-r_4^2\right) \sin \left(\kappa_2 \phi -\frac{\pi }{2}\right)}}\,,
\end{equation}
where $\kappa_2=\frac{2}{\ell L } \sqrt{\left(r_S^2-r_4^2\right) \left(r_S^2-r_F^2\right)}$. In Fig. \ref{SRNfig:anguloalfa}, we plot the above solution and we can see a special second kind trajectory with a negative precession.

\subsubsection{Critical trajectories.}

\begin{figure}[H]
	\begin{center}
		\includegraphics[width=3cm]{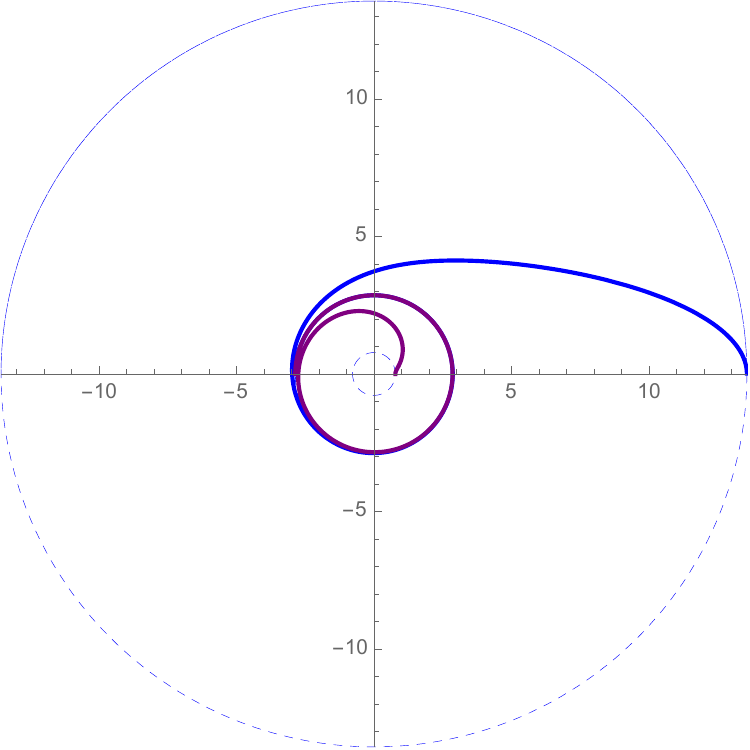}
	\end{center}
	\caption{ The behaviour of critical trajectories with  $M=1$, $\ell=10$, $Q=1.2$, and $L=6$. Blue line for CFK and purple line for CSK trajectories with $E=E_U=1.836$, and $r_u=2.861$.
	The small circle represents to the return point $d_0$ inside the Cauchy horizon, the medium circle represents to $r_u$, and the bigger circle represents to $R_0$.   }
	\label{fig:critical}
\end{figure}

This kind of motion is indeed ramified into two cases; critical trajectories of the first kind (CFK) and those of the second kind (CSK) where the particles start from an initial point $d_i$ at the vicinity of $r_i$ and then tend to this radius by spiraling. We obtain the following equations of motion for the aforementioned trajectories:

\begin{equation}
r_{CFK}(\phi)=\sqrt{r_u^2-\frac{2 \left(r_u^2-d_0^2\right) \left(R_0^2-r_u^2\right)}{\left(d_0^2+R_0^2-2 r_u^2\right)-\left(R_0^2-d_0^2\right) \cosh (\kappa_c \phi )}}\,,
\label{criticas1RN}
\end{equation}
\begin{equation}
r_{CSK}(\phi)=\sqrt{r_u^2-\frac{2 \left(r_u^2-d_0^2\right) \left(R_0^2-r_u^2\right)}{\left(R_0^2-d_0^2\right) \cosh (k_c \phi )+\left(d_0^2+R_0^2-2 r_u^2\right)}}\,,
\label{criticas2RN}
\end{equation}
where  $\kappa_c=\frac{2 \sqrt{\left(r_u^2-d_0^2\right) \left(R_0^2-r_u^2\right)}}{\ell L}$,
$R_0$ is $\left. r_{A} \right|_{E_U}$, and $d_0$ is  
$\left. r_{4} \right|_{E_U}$. In Fig. \ref{fig:critical}, we show the behaviour of the CFK (blue line) and CSK (purple line) trajectories, given by Eq. (\ref{criticas1RN}) and Eq. (\ref{criticas2RN}). 

\subsubsection{Motion with $L=0$} 

\begin{figure}[H]
	\begin{center}
\includegraphics[width=50mm]{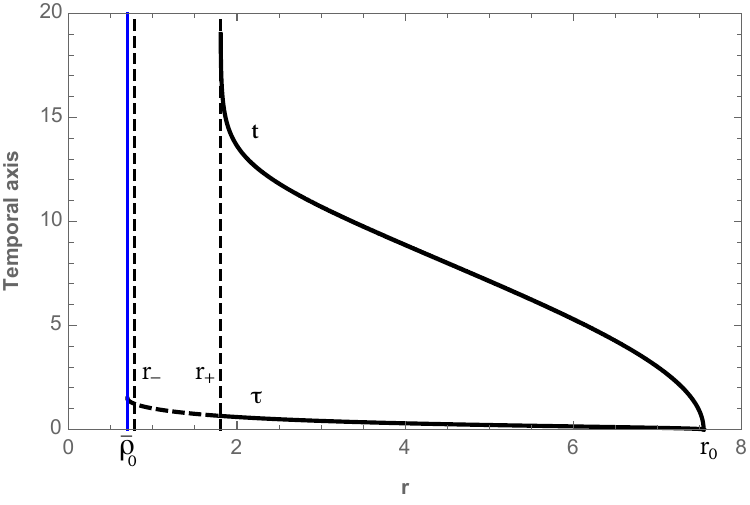}
	\end{center}
	\caption{Proper and coordinate time for the motion of test particles with $L=0$.
	Here we have used the values $E\approx 1.225$, $Q=1.2$, $M=1$, $\ell=10$, and $r_0\approx 7.547$. Blue line corresponds to the return point ($\bar{\rho}_0=0.699$) inside the Cauchy horizon ($r_-\approx 0.783$), the dashed line corresponds to the Cauchy and event horizon ($r_+\approx 1.805$).}
	\label{f2RN}
\end{figure}

As mentioned, for the motion of particles with $L=0$ the effective potential is $V^2=f(r)$. Consequently, the equations governing this kind of motion are (\ref{mr.1}) and (\ref{mr.2}).
Thus,
by assuming that particles are placed at $r=r_0$
when $t=\tau=0$, so, the turning point $r_0$ is located at 
\begin{equation}
r_{0}=\sqrt{Z_0 \cos \Theta_0 -\ell^2(1-E^2)/3}\,,
\end{equation}
where $Z_0=2 \sqrt{\alpha_0/3}$ and
$\Theta_0=\frac{1}{3} \arccos \left(- \frac{3 \sqrt{3} \beta_0}{2 \alpha_0^{3/2}}\right)$, with 
\begin{eqnarray}
  	\label{mr54}
&&\alpha_0 ={\ell^4(1-E^2)^2\over 3}+4\,M^2\,\ell^2\,, \\
&&\beta_0={2\,\ell^6(1-E^2)^3\over 27}+{4\,M^2\,\ell^4(1-E^2)\over 3}+Q^4\,\ell^2\,.
  \end{eqnarray}
Therefore, by replacing Eq. (\ref{mr.1}) into Eq. (\ref{mr.2}), we obtain the proper time ($\tau$) for the motion of massive particles with $L=0$ in terms of the radial coordinate ($r$), resulting
\begin{equation}\label{t4RN}
\tau \left( r\right) ={r_0^3\over 8\,Q^2} \left[F_0(r)-F_0(r_0)\right]\,,
\end{equation}%
and the coordinate time ($t$) as function of $r$ yields
\begin{equation}\label{t5RN}
t\left( r\right) =A_0\,\sum _{i=1}^{3}A_i\,\left[F_i(r)-F_i(r_0) \right]\,,
\end{equation}
where the functions $F_j(r)$ ($j=0,1,2,3$) are given explicitly by
\begin{eqnarray}
  	\label{F0}
&&F_j(r)={\zeta\left(\Omega_j \right)\wp^{-1}\left[U(r)\right]+
\ln\left|\frac{\sigma\left[\wp^{-1}\left[U(r)\right]-
\Omega_j\right]}
{\sigma\left[\wp^{-1}\left[U(r)\right]+
\Omega_j\right]}\right|\over \wp^{\,\prime}\left(\Omega_j\right)}
\,,\\
&&U(r)={M^2\,r_0^2\over 3Q^4}-{r_0^2\over 4r^2}\,,
  \end{eqnarray}
where the invariants are given by 
\begin{eqnarray}
\label{inv2}
&&g_2={4M^4r_0^4\over 3\,Q^8}-{(1-E^2)r_0^4\over 4\,Q^4}\,,\\
&&g_3={M^2(1-E^2)r_0^6\over 12\,Q^8}-{8M^6r_0^6\over 27\,Q^{12}}+{r_0^6\over 16\,Q^4\,\ell^2}\,.
\end{eqnarray}
and
\begin{equation}
    \Omega_0=\wp^{-1}\left({M^2r_0^2\over 3Q^4}\right)\,,\,\,
\Omega_1=\wp^{-1}\left({M^2r_0^2\over 3Q^4}-{r_0^2\over 4\,r^2_+}\right)\,,
\end{equation}
\begin{equation}
    \Omega_2=\wp^{-1}\left({M^2r_0^2\over 3Q^4}-{r_0^2\over 4\,r^2_-}\right)\,,\,\,
\Omega_3=\wp^{-1}\left({M^2r_0^2\over 3Q^4}+{r_0^2\over 4\,r^2_3}\right)\,,
\end{equation}
with the corresponding constants
\begin{equation}
A_0={E \ell^2r_0^7\over 8Q^2r_+^2r_-^2r_3^2}\,,\,\,
A_1={r_+^4r_-^2r_3^2\over r_0^4(r_+^2-r_-^2)(r_+^2+r_3^2)}\,, 
\end{equation}
\begin{eqnarray}
\label{inv2}
&&A_2=-{r_+^2r_-^4r_3^2\over r_0^4(r_+^2-r_-^2)(r_-^2+r_3^2)}\,,\\
&&A_3=-{r_+^2r_-^2r_3^4\over r_0^4(r_-^2+r_3^2)(r_+^2+r_3^2)}\,,
\end{eqnarray}  
where $r_{3}\equiv|\sqrt{x_2-\ell^2/3}|$. Now, in order to visualize the behaviour of the proper and coordinate time, we plot in Fig \ref{f2RN}, their behaviour via the solution (\ref{t4RN}), and (\ref{t5RN}), respectively.
We can observe that the particles 
cross the event horizon  in a finite proper time, but an external observer will see that the particles take an infinite (coordinate) time to do it, the same behaviour was observed for the uncharged background.

\section{Final remarks}
\label{conclusion}

Through this work we have considered the motion of neutral particles in the background of five-dimensional Schwarzschild Anti de Sitter (SAdS) and Reissner Nordstr\"om Anti de Sitter (RNAdS) black holes, and we established the time-like structure geodesic in order to complete the geodesics structure of this spacetime, and also to study the effect of the black hole charge in the geodesics. As we have seen, the time-like geodesics structure have been determined analytically and also the geodesics have been plotted in order to see their behaviour.\\ 

Concerning to the horizons, the five-dimensional 
SAdS  black hole present only the event horizon, while that RNAdS black hole present the event horizon and the Cauchy horizon. At the level of the effective potential, the trajectories are qualitatively similar for fixed values of the mass, cosmological constant, angular momentum and energy. However, for the charged case there is a return point located at $r<r_-$, which implies that the trajectory do not reach the singularity $r=0$. Specifically, for the circular orbits, we have shown that the effect of the charge is to increase the energy necessary in order to obtain an unstable circular orbit, but the radius of such orbit decreases. With respect to the stable circular orbit, the effect of the charge  is to increase the energy necessary in order to obtain the stable circular orbit, but the radius of such orbit increases. Also, the effect of the black hole charge is to increase the affine and coordinate period. For the planetary orbits, the effect of the black hole charge is to increase the periastro distance and to decrease the apoastro distance. Also, both spacetimes allow planetary orbits with negative, null and positive precession. But, the black hole charge increases the energy value for a null precession. For the other relativistic orbits, as second kind and critical trajectories, the black hole charge implies the existence of a return point between the singularity and the Cauchy horizon. For the motion with $L=0$ in the Schwarzschild AdS  black hole the particles cross the event horizon  in a finite proper time and reach the singularity, but for an external observer the particles take an infinite coordinate time to reach the event horizon. However, the difference for charged spacetime is that particles that cross the event horizon in finite proper time cannot reach the singularity. \\

Could be interesting to consider the motion of charged particles in order to analyze the influence of the charge of the particles on the orbits, we hope to address it, in a forthcoming work.\\

\section*{Acknowlegements}

This work is partially supported by ANID Chile through FONDECYT Grant Nº 1220871  (P.A.G., and Y. V.). J.R.V. was partially supported by Centro de Astrof\'isica de Valpara\'iso.

\end{document}